\documentstyle[12pt,epsfig]{article}
\def\beq{\begin{equation}}
\def\eeq{\end{equation}}
\def\bea{\begin{eqnarray}}
\def\eea{\end{eqnarray}}
\def\bq{\begin{quote}}
\def\eq{\end{quote}}

\def\gappeq{\mathrel{\rlap {\raise.5ex\hbox{$>$}}
{\lower.5ex\hbox{$\sim$}}}}
\def\lappeq{\mathrel{\rlap{\raise.5ex\hbox{$<$}}
{\lower.5ex\hbox{$\sim$}}}}
\def\Toprel#1\over#2{\mathrel{\mathop{#2}\limits^{#1}}}

\begin{document}
\pagestyle{empty}
\begin{flushright}
{CERN-TH/2001--329}\\
{PAR/LPTHE/01--39}\\
\end{flushright}
\vspace*{5mm}
\begin{center}
{\bf  Strangeness enhancement  and Energy dependence \\
 in\\
 Heavy Ion Collisions}  \\
\vspace*{1cm}
{\bf K. Redlich$^{a,*)}$ and  A. Tounsi$^{b)}$ }\\
 $^a$
Theoretical Physics Division, CERN, CH-1211 Geneva 23, Switzerland  \\
 $^b$
  Laboratoire de Physique Th\'eorique et Hautes Energies,\\
 Universit\'e Paris 7, F-75251 Cedex 05, France

 \vspace*{2cm} {\bf ABSTRACT} \\
  \end{center} \vspace*{1mm}
\noindent The canonical statistical model analysis of strange and
multistrange hadron production in central A--A relative to
p--p/p--A collisions is presented over the  energy range from
$\sqrt s=8.73$ GeV up to $\sqrt s =130$ GeV. It is shown that the
relative enhancement of strange particle yields from p--p/p--A to
A--A collisions substantially increases with decreasing collision
energy. It is  largest at $\sqrt s= 8.7$ GeV, where the
enhancement of $\Omega,\Xi$ and $\Lambda$  is of the order of 100,
20 and 3, respectively.  In terms of the model these results are
due to the canonical suppression of particle thermal phase space
at lower energies, which increases with the strangeness content of
the particle and with decreasing  size of the collision fireball.
The comparison of the model with existing data on energy
dependence of the kaon/pion ratio  is also discussed.

\vspace*{4.5cm} \noindent \rule[.1in]{16.5cm}{.002in}

\noindent $^{*)}$ Permanent address: Institute of Theoretical
Physics, University of Wroc\l aw, PL-50204 Wroc\l aw, Poland:
redlich@rose.ift.uni.wroc.pl \vspace*{0.5cm}

\begin{flushleft} CERN-TH/2001-329\\
November 2001
\end{flushleft}
\vfill\eject

\setcounter{page}{1}
\pagestyle{plain}

\section{Introduction}

The production of strange particles  is extensively studied in
heavy ion experiments in a very broad energy range from GSI/SIS to
BNL/AGS  and from CERN/SPS  up to BNL/RHIC \cite{rev1}. The amount
of data available for $K^+$ and $K^-$ yields already  allows  a
detailed analysis of the kaon excitation function in A--A
collisions \cite{stoc,ogilvie,blume} as well as an analysis of the
relative enhancement of kaon production in A--A with respect to
p--p collisions \cite{ogilvie,cleymans1,gaz1}. Of particular
interest is here the behaviour of $K^+$ yield with energy, as this
is the most abundantly produced particle with non--vanishing
strangeness. The kaon excitation function at midrapidity, as the
$K^+/\pi^+$ ratio, is a very abruptly increasing function of the
collision energy between SIS up to top AGS. At higher energies it
reaches  a broad maximum between 10 AGeV  and 40 AGeV
\cite{cleymans1} and, gradual decrease up   to RHIC energy
\cite{harris,star1}. In the microscopic transport models the
increase of the kaon yield with collision energy is qualitatively
expected, owing to a change in the production mechanism from
associated to direct kaon emission. However, the hadronic cascade
transport models are as yet  not providing the quantitative
explanation of the experimental data. The Hadron String Dynamics
(HSD) model \cite{cassing} severely underpredicts the top AGS
results on $K^+/\pi^+$ ratio, while RQMD \cite{ogilvie}
overpredicts the yield at lower energies and gives too small yield
at the SPS. The Statistical model SM, on the other hand, provides
quite a satisfactory description of the kaon excitation function
and of the $K/\pi$ midrapidity ratio in the whole energy range
from SIS up to RHIC \cite{cleymans2,redqm}. A detailed analysis of
the experimental data at SIS \cite{cleymans2}, AGS \cite{pbm1,bc},
SPS \cite{bc,pbm2} and recently also at RHIC \cite{pbm3,nuxu,pol}
has shown that hadronic yields and their ratios resemble those of
a population in chemical equilibrium. Most particle multiplicities
measured in nucleus--nucleus collisions are well consistent with
the thermal model predictions \cite{bc,pbm2,pbm3,mark}. A similar
analysis of particle yields in hadron--hadron collisions \cite{b1}
also shows the  consistency of the statistical model with the
data.

Recently, it was argued \cite{hamieh} that  SM can also describes
a basic future of strangeness enhancement from p--A to A--A
collisions measured by WA97 Collaboration \cite{wa97}: the
enhancement pattern and enhancement saturation for a large number
of participants  $N_{\rm part}$. The enhancement of (multi)strange
baryons is of particular interest, since  it was conjectured to be
a signal for quark--gluon plasma formation in heavy ion collisions
\cite{rafelski}. The centrality dependence of the (multi)strange
baryon enhancement, increasing with the strangeness content of the
particle, was shown  to appear in the SM as a consequence of the
canonical suppression of the particle phase space in a small
system. Satisfactory agreement of the SM with data confirmed that
SM provides a tool to make a prediction for hadron production in
A--A collisions.

In this paper we apply (SM) formulated in the canonical ensemble
\cite{hagedorn,newpbm} to study the energy dependence of the
(multi)strange baryon enhancement from p--p to A--A collisions. We
first show that the observed  relative enhancement of the
$K^+/\pi^+$ ratio in A--A to p--p collisions from top AGS up to
top SPS energy can be quantitatively described by the SM. We then
make a prediction for relative (multi)strange baryon enhancement
at $\sqrt s = 8.73, 12.3, 17.3$ GeV at SPS and $\sqrt s=130$ GeV
at RHIC energies. We show that at RHIC the enhancement of
(multi)strange baryons is expected to be comparable with the one
measured at the top SPS energy; however, it could be almost an
order of magnitude larger at lowest ($\sqrt s=8.73$ GeV) SPS
energy.


\section{Statistical description of strangeness enhancement}

Within  the statistical approach, hadron production  is commonly
described  by using the grand canonical GC partition function,
where the charge conservation is controlled by the related
chemical potential. In this description a net value of a given
U(1) charge is conserved on the average. The GC approach can be
valid only if the total number of particles carrying a quantum
number related with this symmetry is very large. In the opposite
limit of  small particle multiplicities, conservation laws must be
implemented exactly and locally, i.e. the canonical C ensemble for
conservation laws must be used \cite{hagedorn,ko1,ko2}. The local
conservation of quantum numbers in the canonical approach severely
reduces the thermal  phase space available for particle
production. The exact charge conservation is therefore of crucial
importance in the description of particle yields in
proton--induced processes \cite{b1,hamieh}, in $e^+$--$e^-$
\cite{b1}, as well as in peripheral heavy ion collisions
\cite{cs}.

In the present analysis of the energy dependence of the
strangeness enhancement, we adopt the statistical model, which has
 previously been used to  analyse  hadron productions  in heavy ion
collisions \cite{hamieh,newpbm}. This model is formulated in the
canonical ensemble with respect to strangeness conservation,
whereas  baryon number  conservation is controlled by  the
chemical potential. Within this framework particle multiplicity
ratios depend only on two thermal  parameters: the temperature $T$
and the baryon chemical potential $\mu_B$, as well as the
correlation volume, which is assumed to scale with the number of
 participating projectile nucleons.

The canonical model description of strangeness conservation
including  multistrange particle contributions was  described in
detail in \cite{hamieh,newpbm,cs}. We quote here the final results
\cite{newpbm} for particle density, which are relevant to our
discussion of the energy dependence of the strangeness
enhancement.

 The number density $n_i^s$ of
particle $i$ carrying strangeness $s$ is found to be
\begin{equation}
n_{i}^s=\frac{Z^1_{i}}{Z^C_{S=0}}
\sum_{n=-\infty}^{\infty}\sum_{p=-\infty}^{\infty} a_{3}^{p}
a_{2}^{n}
 a_{1}^{{-2n-3p- s}} I_n(x_2) I_p(x_3) I_{-2n-3p- s}(x_1)
 , \label{eq1}
\end{equation}
where
\begin{equation}
Z^C_{S=0}= \sum_{n=-\infty}^{\infty}\sum_{p=-\infty}^{\infty}
a_{3}^{p} a_{2}^{n} a_{1}^{{-2n-3p}} I_n(x_2) I_p(x_3)
I_{-2n-3p}(x_1)
  \label{eq2}
\end{equation}
is the canonical partition function for a system with total
stangeness $S=0$. The variables $a_i$ and $ x_i $ are defined by
 $a_i= \sqrt{{S_i}/{S_{-i}}}$, $ x_i = 2\sqrt{S_iS_{-i}}$,
and $S_n= V\sum_k Z_k^1$ is  the sum  over all particles and
resonances carrying strangeness $n$. For a particle of  mass
$m_k$, with degeneracy factor $g_k$, carrying  baryon number $B_k$
with the corresponding chemical potentials $\mu_B$,
 the one-particle partition function
 in Boltzmann approximation reads
\begin{equation}
 Z_k^1\equiv {{g_k}\over {2\pi^2}}
m_k^2TK_2\left({{m_k}\over T}\right)\exp (B_k\mu_B) ;\label{e:zk1}
\end{equation}
here  $I_i$ and $K_2$ are the modified Bessel functions.

It can be   shown  that, in the limit of large $x_1$, that is for
large strangeness multiplicity, the above formula coincides with
the grand canonical result.\footnote{ This can be easily  seen for
charges and baryons neutral system by making a large argument
expansion of the Bessel functions in Eq.~(\ref{eq1}).} In the
opposite limit, that of small $x_1$, the equilibrium density of
(multi)strange baryons is strongly suppressed with respect  to
their grand canonical value \cite{hamieh}.

 To study the energy and centrality dependence of (multi)strange
particle yields in terms of the above model, one  needs to
establishes first the variation  of  thermal parameters with
energy and centrality. From a previous analysis \cite{b3} one
knows that  temperature to a good approximation is only a function
of the collision energy and is independent from the number of
participating nucleons. For fully integrated,  $4\pi$ particle
yields, one can also assume that the  baryon chemical potential is
weakly changing with centrality. The above assumption could be
partly justified  by data since for $\sqrt s =8.73$ GeV the total
number of pions per participant in central Pb--Pb ($\langle
\pi\rangle /\langle N_{\rm part}\rangle $=2.5) and in  p--p
($\langle \pi\rangle /\langle N_{\rm part}\rangle \sim2.62$)
almost coincide \cite{gr} (see also Fig. 1). In the SM  and with
the same temperature, the above can only be valid if the baryon
chemical potential is weakly centrality--dependent. For larger
collision energies, such as  at RHIC, $\langle \pi\rangle /\langle
N_{\rm part}\rangle $ is seen in Fig. 1 to increase from p--p to
central Au--Au collisions. Thus, here $\mu_B$ is decreasing with
increasing centrality. With a freeze--out temperature $T\sim 175$
MeV and $\mu_B\sim 50$ MeV required to reproduce the RHIC data
\cite{newpbm}, this would correspond to an increase of  $\mu_B$
from 50 MeV at most central Au--Au collisions to 70 MeV in p--p
collisions. This change, however, will only weakly influence the
relative enhancement of (multi)strange particles. We thus assume,
for simplicity,  that the baryon chemical potential is
centrality--independent.\footnote{ In general in p--p collisions
the baryon number should also be treated canonically \cite{bc}.
However, since here the baryon number $B=2$ the canonical
suppression is much less effective than in the case of
strangeness. The canonical corrections due to baryon number should
not exceed $30\%$ \cite{gazdzickil}.}
\begin{figure}[htb]
\begin{minipage}[t]{80mm}
\includegraphics[width=17.pc, height=19.3pc]{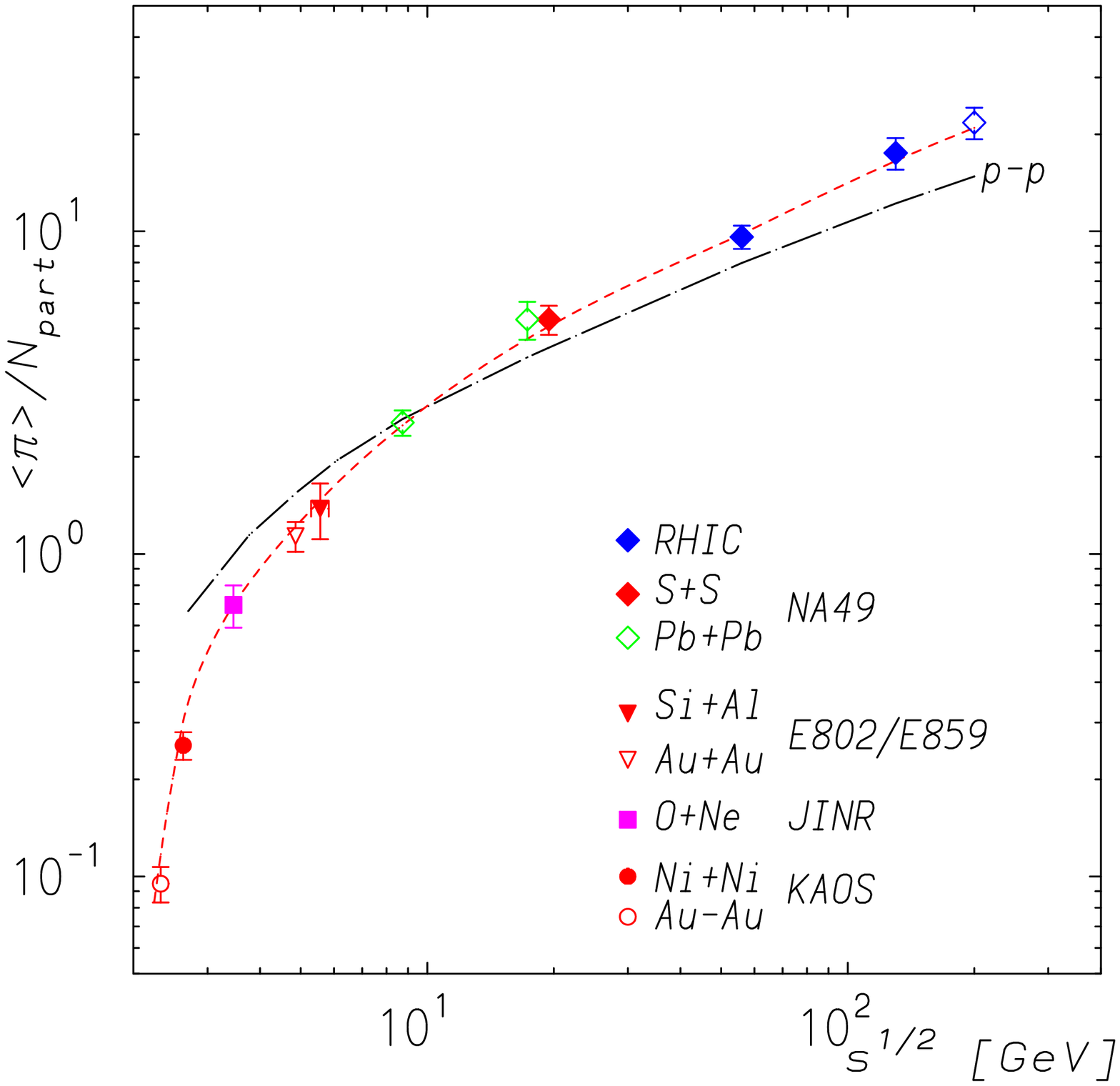}\\
\vskip -1. true cm \caption{Total number of pions  per participant
($\langle \pi\rangle /N_{\rm part}$), 4$\pi$ ratio versus energy.
The data at lower energies in A--A  as well as in p--p collisions
are from \cite{gr}. The RHIC results are from \cite{rhic}. The
short-dashed and dashed lines represent the fit to the data. }
\end{minipage}
\hspace{\fill}
\begin{minipage}[t]{75mm}
\includegraphics[width=17.pc, height=19.2pc]{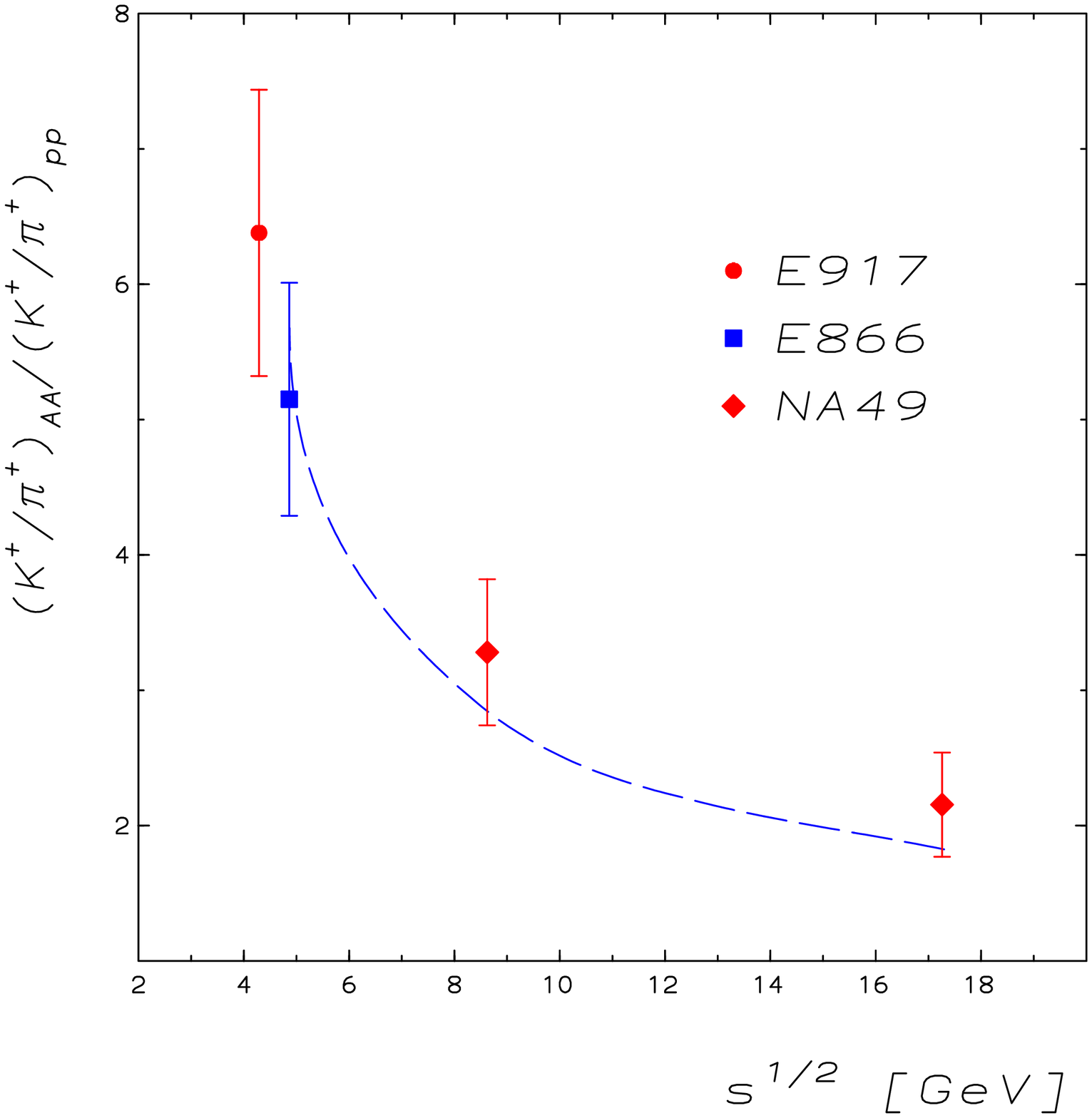}\\
\vskip -1. true cm \caption{$K^+/\pi^+$ ratio  in A+A relative to
p+p collisions. For the compilation of data, see
\cite{ogilvie,blume,redqm}. The dashed line represents the
statistical model results. \hfill}\label{kpi}
\end{minipage}
\end{figure}
Thermal parameters are, however, very sensitive to collision
energy. At the AGS \cite{pbm1} and SPS  \cite{pbm2} we use the
chemical freeze--out parameters as obtained from a detailed
analysis of the experimental data. In particular at
$\sqrt{s}=17.3$ GeV, $T \sim 168$ MeV and $\mu_B \sim 266$ MeV. At
RHIC we adopt the values from \cite{newpbm}. At 40 AGeV the
available data are still not sufficient to make a precise
determination of freeze--out parameters \cite{bs}. To estimate
these values we use the relation between $T$ and $\mu_B$ as
determined by the unified freeze--out conditions of fixed energy
per particle \cite{prl} and the measured result on $\langle
\pi\rangle /N_{\rm part} =2.55\pm 0.15$ for Pb--Pb collisions at
40 AGeV \cite{blume}. The Pb--Pb data at 80 AGeV for $\langle
\pi\rangle /N_{\rm part}$ are still not known. Here we use instead
the extrapolated to 80 AGeV value from Fig. 1, leading to $\langle
\pi\rangle /N_{\rm part} \sim 3.4 $. Thermal parameters for 40 and
80 AGeV extracted in this way are: $T=145\pm 5$ MeV, $\mu_B=370\pm
20$ MeV and $T\sim 152$ MeV, $\mu_B\sim 280$ MeV correspondingly.
Finally, the volume parameter appearing in Eq.~(\ref{eq1}) is
assumed to be proportional to the number of projectile
participants, $V\simeq V_0N_{\rm part}/2$ where $V_0\simeq 7$
fm$^3$ is taken as the volume of the nucleon.


 In Fig. \ref{kpi} we show the compilation of the data on
the $K^+/\pi^+$ ratio in A--A relative to p--p collisions
\cite{ogilvie,blume}. This double ratio could be referred to as
strangeness enhancement factor. The enhancement is seen in the
data to be the largest at the smallest beam energy and is
decreasing towards higher energy. The line is a smooth
interpolation between the canonical  model results for $\sqrt s
=17.3,12.3,8.73, 5.56$ GeV, obtained with the parameters described
above.


The enhancement seen in Fig. \ref{kpi} is entirely due to the
suppression of the $K^+/\pi^+$ ratio in p--p collisions with
decreasing energy and not due to a dilution of this ratio by
excess pions in the A--A system. The $K^+/\pi^+$ ratio is known
experimentally not to vary within $30\%$  in the energy range from
$\sqrt s\sim 5$ GeV at AGS up to $\sqrt s=130$ GeV at RHIC
\cite{ogilvie,blume,redqm}. This behaviour is also described by
the statistical model as a consequence of particular distributions
of thermal parameters with collision energy
\cite{newpbm}.\footnote{ We have to point out that the NA49 data
in the full phase space shows  a depletion in the $K^+/\pi^+$
ratio from 40 A GeV to 160 A GeV. This behaviour  was previously
conjectured as a possible signal of deconfinement \cite{marekr}.}

\begin{figure}[htb]
\vskip -1.cm
\begin{minipage}[t]{75mm}
\includegraphics[width=21.5pc, height=19.5pc]{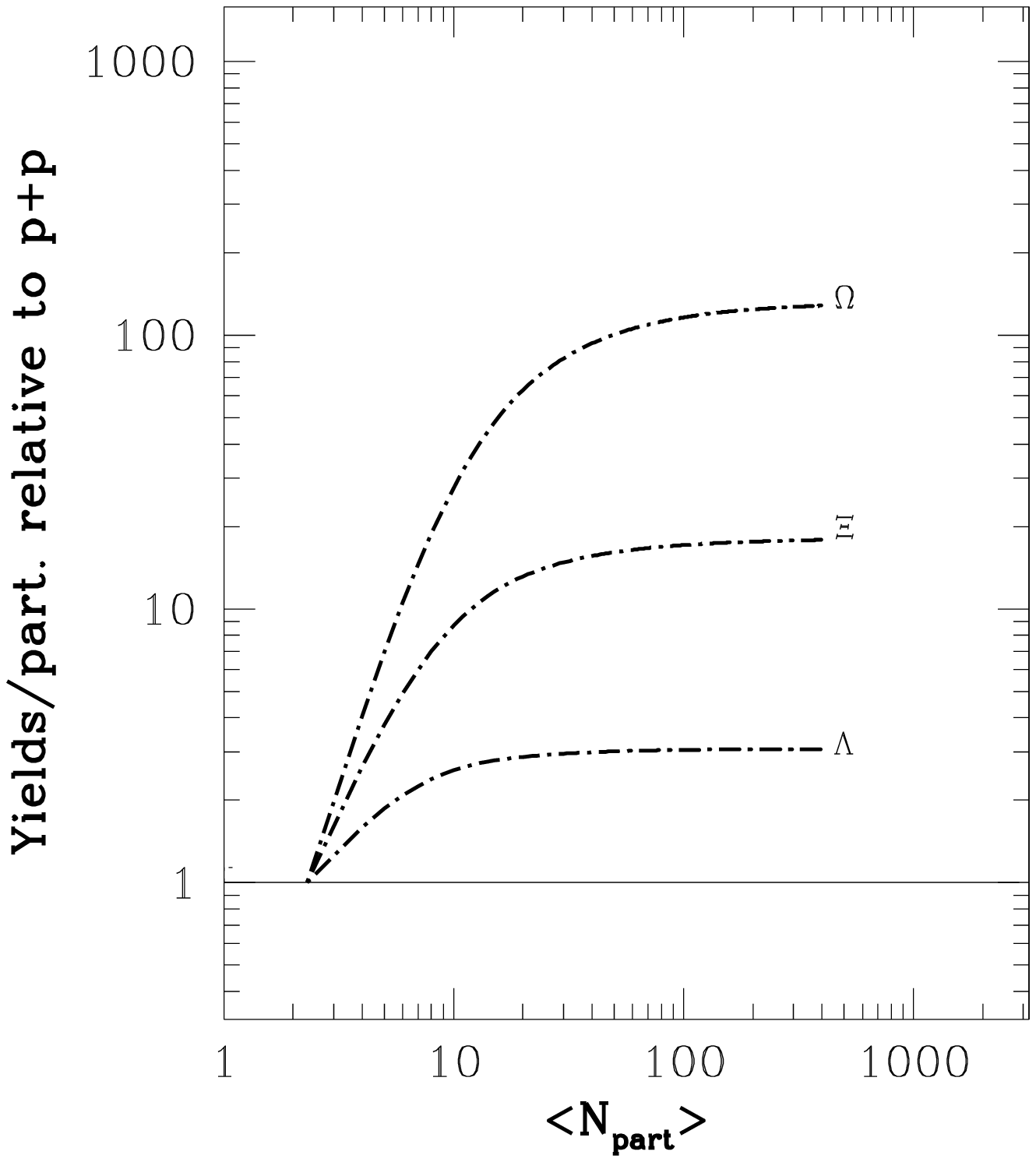}
\vskip -0.5cm \caption{Centrality dependence of the relative
enhancement
 of particle yields/participant in central Pb--Pb to p--p collisions
at fixed energy $\sqrt{s}=8.73$ GeV. \hfill}\label{oxl}
\end{minipage}
\hspace{\fill}
\begin{minipage}[t]{74mm}
\includegraphics[width=21.5pc, height=19.5pc]{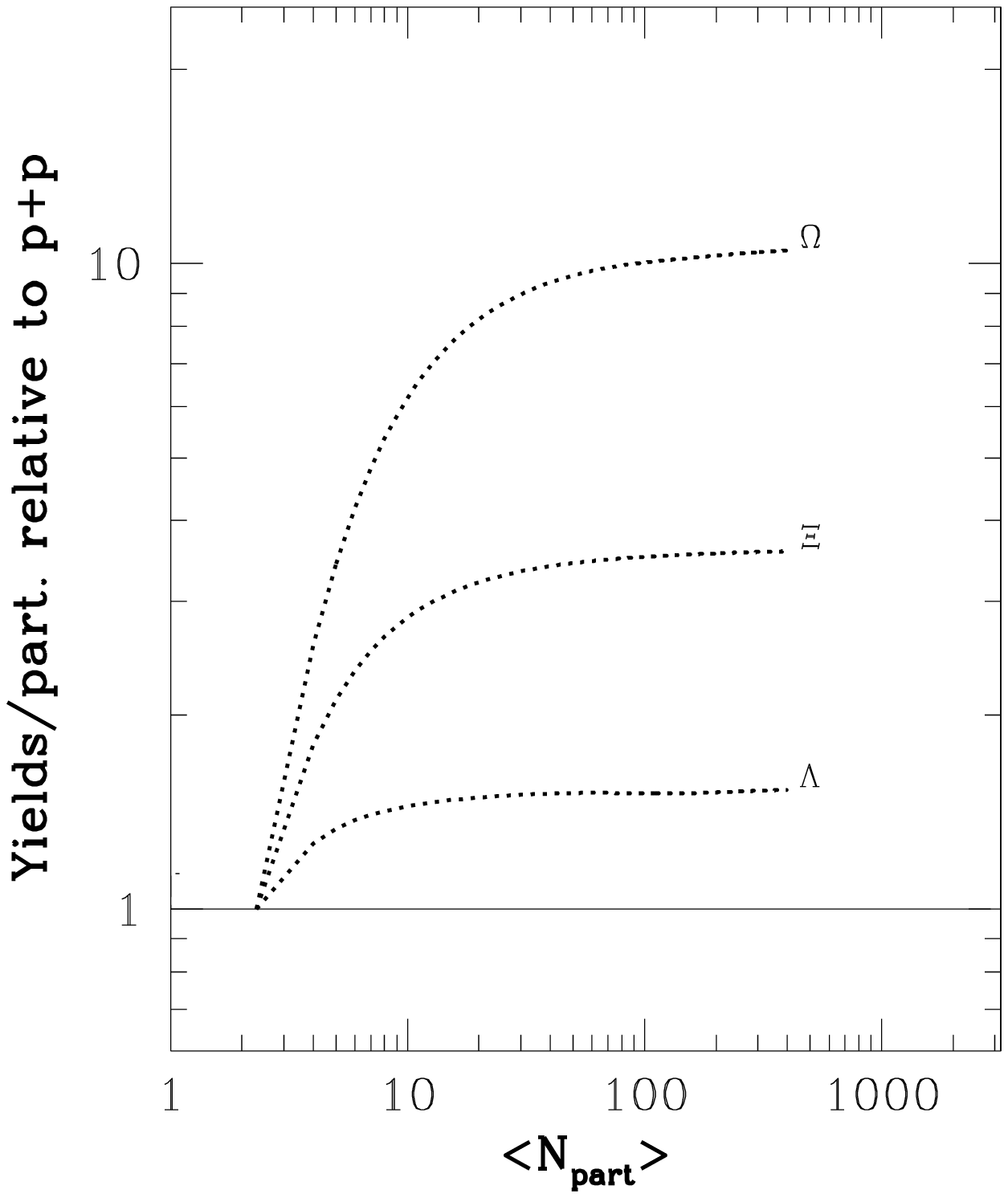}
\vskip -0.5cm \caption{ Centrality dependence of the relative
enhancement of particle yields/participant in central Pb--Pb to
p--p collisions at fixed energy $\sqrt{s}=130$ GeV. \hfill}
\label{oxlr}
\end{minipage}
\end{figure}
\begin{figure}[t]
\vskip -1.cm
\hspace{1cm}
\begin{minipage}[t]{80mm}
\includegraphics[width=21.5pc, height=19.5pc]{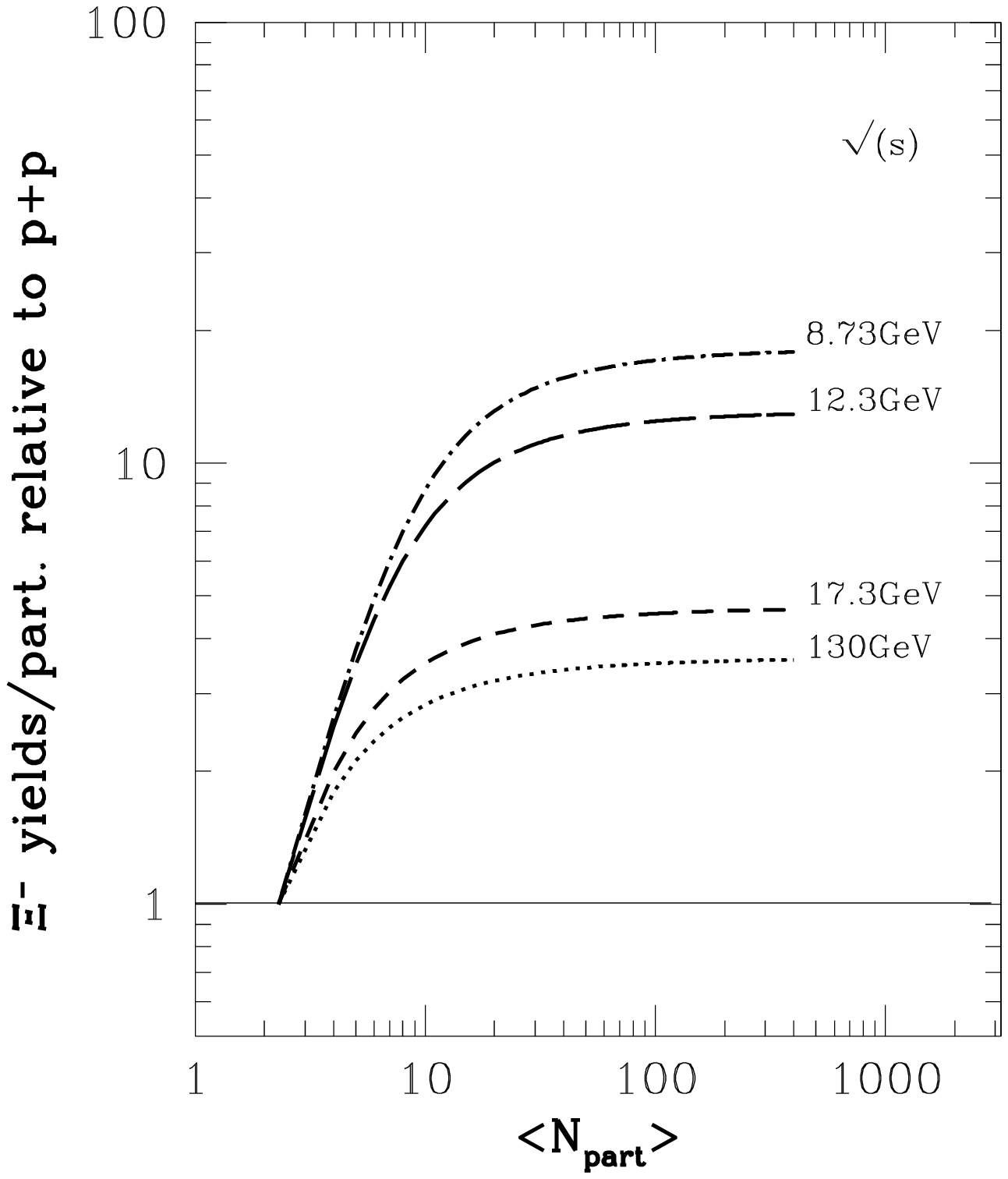}
\vskip -0.5cm \caption{ Centrality dependence of relative
enhancement of $\Xi^-$ yields/participant in central Pb--Pb  to
p--A reactions at different collision energies. \hfill}\label{xi}
\end{minipage}
\hspace{\fill}
\begin{minipage}[t]{76mm}
\includegraphics[width=21.5pc, height=19.5pc]{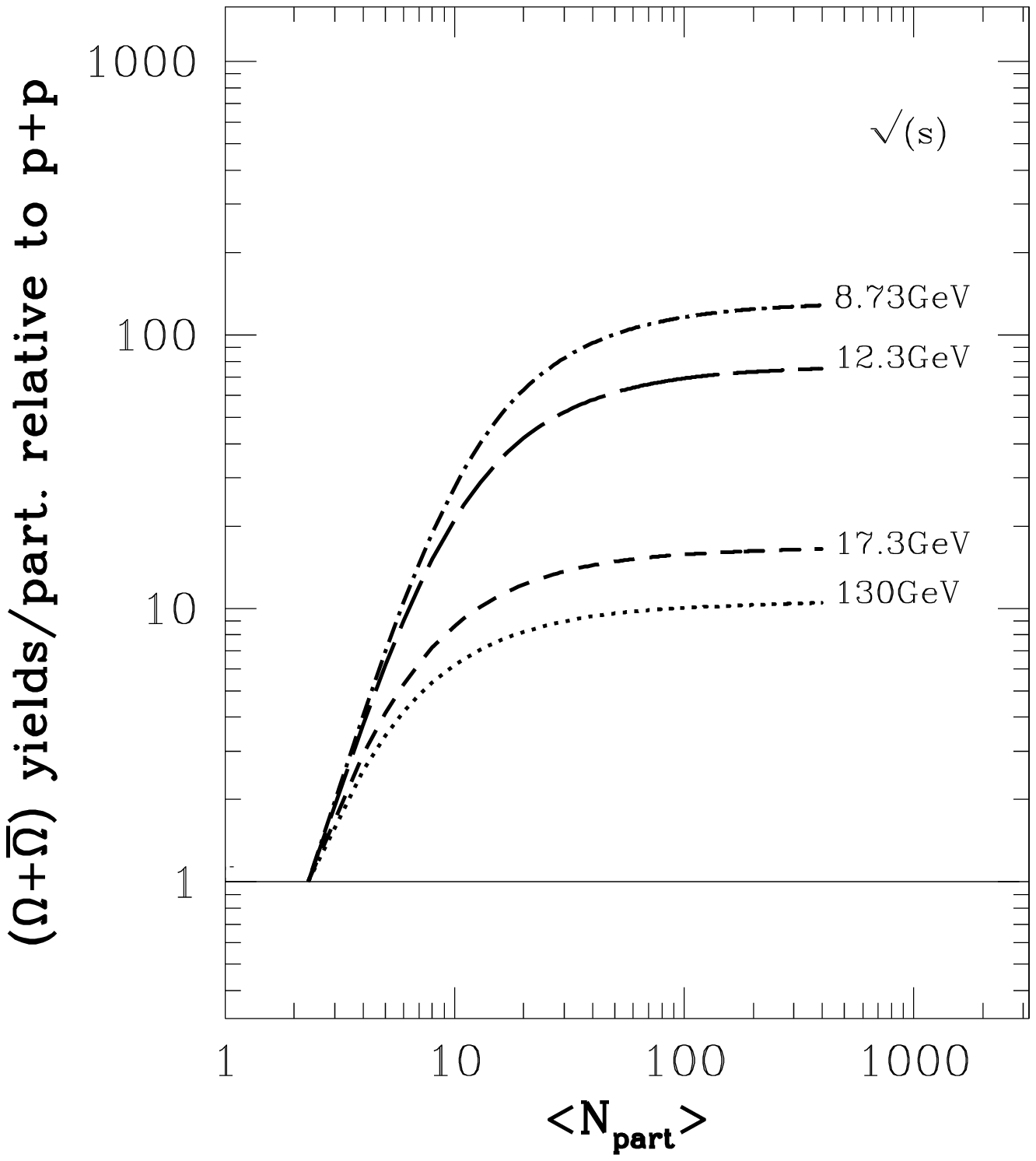}
\vskip -0.5cm \caption{Centrality dependence of relative
enhancement of $({\Omega}+\bar{\Omega})$ yields/participant in
central Pb--Pb to p--p reactions at different collision energies.
\hfill} \label{omega}
\end{minipage}
\end{figure}

The results in Fig. \ref{kpi} show that in terms of the SM the
energy dependence of the strangeness enhancement could be related
with the suppression of the thermal phase space available for
strangeness production in p--p collisions. Having in mind the
above agreement, we apply this model to analyse  the enhancement
for multistrange baryons.  Two relevant questions are here of
particular interest: (i)
 Is the enhancement
pattern observed by WA97 a characteristic feature of the top SPS
energy or can
 it also be seen at lower energies ?  (ii) Is the enhancement for
multistrange baryons  a decreasing or increasing function of
energy? In Fig. \ref{oxl} and Fig. \ref{oxlr}
 we show the results on relative (multi)strange
baryon enhancement from p--p to Pb--Pb collisions  at $\sqrt
s=8.73$ GeV  and at $\sqrt s=130$ GeV, respectively. It is clear
that   the same enhancement pattern as at the SPS is expected in
the SM model to appear for all relevant energies.

\begin{figure}
\hspace*{4cm}
\includegraphics[width=21.5pc, height=20.5pc]{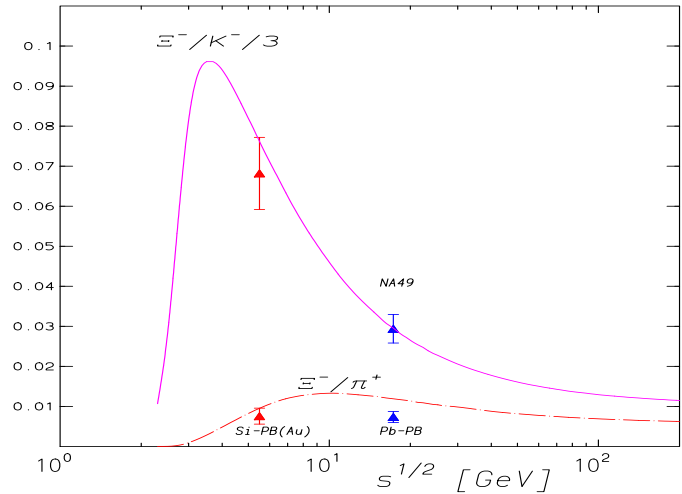}
\vskip -1.3cm \caption{{ Ratios of $\Xi^-/\pi^+$ and $\Xi^-/K^-$
as a function of collision energy. The data at the SPS are fully
integrated NA49 results. The corresponding ratio at the top AGS
was obtained from E810 results on $\Xi^-$ measured in Si--Pb
collisions in the rapidity interval $1.4\l y\l 2.9$ \cite{e810}
normalized to the full phase space values of $\pi^+$ and $K^-$
yield obtained in Si-Au collisions by E802 \cite{e866}. }}
\label{last}
\end{figure}

 To see the  dependence of the
strength of the enhancement with energy, we show in Fig. \ref{xi}
and Fig. \ref{omega} the relative enhancement of $\Xi$ and
$(\Omega+\bar\Omega )$   for different collision energies. The
enhancement is the largest at lowest energy;  for $\Omega$, it can
 even be larger by a factor of almost 10  at 40 A GeV than observed
at the SPS. This behaviour could  already be partly deduced from
the experimental data. Indeed,  combing the Si--Pb   results for
$\Xi^-$ production obtained by  the E810 Collaboration \cite{e810}
and the E802 value for pion or $K^-$ yield in Si--Au collisions
\cite{e866} at the top AGS energy, one can estimate that
$\Xi^-/\pi^+\sim 0.0076$; this within errors, coincides with the
value of $\Xi^-/\pi^+\sim 0.0074$ obtained by NA49 in Pb--Pb at
$\sqrt s =17.3$ GeV \cite{bna49}, as seen in Fig. \ref{last}. In
p--p collisions the $\Xi^-/\pi^+$ ratio is obviously a strongly
decreasing function of energy.  From the above one could therefore
expect that the relative enhancement of $\Xi^-$ from p--p, p--A to
A--A collisions should be larger at AGS than at SPS energy, which
is seen in Fig. \ref{xi}. In addition the ratio containing only
newly produced  particles such as e.g. $\Xi^-/K^-$, is also seen
in Fig. \ref{last} to be larger at AGS than SPS. The energy
dependence of the $\Xi^- /\pi^+$ and $\Xi^- /K^-$ ratios was
calculated in Fig. \ref{last} along the freeze--out curve
\cite{prl} following the method described in \cite{newpbm}.

 The results
of
 Fig. \ref{xi} and Fig. \ref{omega}
  show a
 saturation of the enhancement, which indicates that the grand
canonical limit was reached. It is clear  from these figures that
this saturation  is shifted towards larger centrality with
decreasing energy.

In p--p collisions local strangeness conservation required that
there be associated production of particles, e.g. $\Xi$ has to
appear together with two other strangeness 1 particles, to
neutralize strangeness. In high energy central A--A collisions,
there are already sufficiently many strange hadrons being produced
for strangeness to be conserved on the average, which
substantially increases thermal particle production. The
associated production and locality of strangeness conservation is
an origin of the suppression of particle thermal phase space,
which increases with the strangeness content of the particle as
well as with decreasing collision energy. As a consequence the
enhancement from p--A to A--A collisions with decreasing energy is
seen in Fig. \ref{xi} and Fig. \ref{omega} to be stronger for
$\Omega$ than for $\Xi$. At RHIC the freeze--out temperature was
found to be consistent, within
  errors, with the
one at the SPS as well as with the critical temperature found in
the Lattice Gauge Theory. As a consequence the enhancement of
(multi)strange baryons at $\sqrt s=130$ GeV RHIC energy is seen in
Fig. 4 to be comparable with that at the  SPS. The change in
baryon  chemical potential from 266 MeV to 50 MeV,  evidently does
not influence the strength of the enhancement. Thus, also an
increase of the energy from 130 to 200 GeV should not change the
above results much. These predictions are in obvious disagreement
with UrQMD \cite{bleicher}. There, the production of strangeness
is very sensitive to the initial conditions. In UrQMD the early
stage multiple scattering may imply an increase of the colour
electric field strength. Consequently, according to the Schwinger
mechanism, this should increase the production of (multi)strange
baryons. Under similar kinematical conditions as at the SPS, the
UrQMD model predicts at RHIC an increase in relative strength of
$\Omega$ yields from p--A to A--A by a factor of 4
\cite{bleicher}.


The strangeness enhancement pattern measured  by the WA97
Collaboration at the SPS \cite{wa97} was predicted as a signal for
quark-gluon plasma formation \cite{rafelski}. In the context of
the considered SM model the enhancement pattern of (multi)strange
baryons should be observed at all SPS energies,  with increasing
strength towards lower beam energy.
 Thus, the results of the above SM make it
clear  that strangeness enhancement and enhancement pattern are
not a {\it unique} signal of deconfinement, since  these  features
are expected to be  there also at energies   where the initial
conditions are very unlikely for deconfinement.

The quantitative results shown in Figs. \ref{oxl}--\ref{omega}
contain some uncertainties. The magnitude of the enhancement is
very sensitive to the temperature taken at a given collision
energy. Changing $T$ by 5 MeV, a typical error on $T$ in a thermal
analysis, can change, for instance, the enhancement of $\Omega$
shown in Fig. \ref{oxl} by a factor
 of 2. The $N_{\rm part}$ dependence of the strange hadron enhancement seen in
Figs. 3--6 was obtained by assuming a linear dependence of a
volume parameter $V$ in Eq.~(\ref{eq1}) with $N_{\rm part}$. In
general,
 $V$ could have a
weaker   dependence with centrality, which could be reflected with
an only moderate  increase of the enhancement, with centrality and
saturation appearing at  larger volume. In addition, including the
variation of the thermal parameters, in particular of the
baryon-chemical potential, with centrality, or extending the model
to a canonical description of baryon number conservation
\cite{b1}, or finally including a possible asymmetry between the
strangeness under saturation factors in p--p and A--A collisions
\cite{bc,b1}, could change our numerical values.
However, independently of these uncertainties, the main results:~~
(i) enhancement decreasing with increasing collision energy, and
(ii) enhancement pattern being  preserved at all SPS energies, are
always valid.



\section{Summary and conclusions}

In conclusion, we have  shown that, in terms of the statistical
model, the  relative enhancement of (multi)strange baryons from
proton--proton or proton--nucleus to nucleus--nucleus collisions
is a decreasing function of the collision energy. Experimentally
this fact   already, had been obtained for  kaon yields and is
shown to be expected for multistrange baryons.  In addition, an
increase of the enhancement with the strangeness content of the
particle is a generic feature of our model,  independent of the
collision energy. On the qualitative level  the only input being
required in the model to make the above predictions is the
information that freeze--out temperature is increasing with
collision  energy and that the chemical potential shows the
opposite dependence. The above conditions are well confirmed by a
very detailed analysis of particle production at different
collision energies.

We have presented the quantitative predictions for a relative
enhancement of  $\Lambda$, $\Xi$  and $\Omega$ yields in the
energy range from $\sqrt s=8.7$ up to $\sqrt s=130$ GeV. We have
discussed the possible uncertainties of the presented  results.
The relative enhancement at RHIC was found to be lower than at the
SPS. This is in contrast with the UrQMD  predictions,
 which are showing the opposite  behaviour  \cite{bleicher} .
The statistical model applied here is also very
unlikely to be capable of explaining  an abrupt change of the
$\bar\Xi /N_{\rm part}$ enhancement recently reported by the NA57
Collaboration \cite{carrer}.
\section{Acknowledgements}
One of us, K.R.,  acknowledges a  partial support from the LPTHE
Universite Paris 7 and  the Polish Committee for Scientific
Research (KBN-2P03B 03018). Interesting discussions with F.
Becattini, P. Braun-Munzinger,  J. Cleymans, A. Keranen, H.
Oeschler, H. Satz and R. Stock are also acknowledged.


\begin{thebibliography}{10}
\bibitem{rhic} The 4$\pi$ results at $\sqrt s=56$ GeV were estimated from
PHOBOS mid-rapidity data from \cite{phobos}, scaled by the same
factor as required in p--p collisions when going from mid-rapidity
to full phase space. The results at $\sqrt s=130$ GeV are from
\cite{brahms1}. The 4$\pi$ results at $\sqrt s=200$ GeV were
estimated by multiplying the $\sqrt s=130$ GeV results by  the
same factor (modulo an increase in rapidity interval) as recently
measured by PHOBOS at mid--rapidity \cite{phobos}.
\end{thebibliography}
\end{document}